# EFFECTS OF PHOTOIONIZATION ON GALAXY FORMATION

Masashi Chiba[1,2] and Biman B. Nath[1]

[1]Max-Planck-Institut für Radioastronomie, Auf dem Hügel 69, D-53121 Bonn, Germany

[2]Astronomical Institute, Tohoku University, Aoba-ku, Sendai 980, Japan

## ABSTRACT

We investigate the effects of the observed UV background radiation on galaxy formation. Photoionization by UV radiation decreases the cooling rate of the gas in halos, so that objects with only large density contrasts can self-shield against the background radiation, thereby allowing the shielded neutral cores to cool and form stars. In the context of the CDM model, we use the criterion that self-shielding is essential for star formation to calculate the mass function of galaxies based on both Press-Schechter and the peaks formalisms. The ionizing UV radiation results in the inhibition of galaxy formation — we show the decrease in the number density of the galaxies quantitatively. We also find that the merging in general is made inefficient by the UV photons through photoionization of the gas in the bigger system into which small objects are incorporated. The latter means that, in a merging dominated region, where the number density at the low-mass end ($M_b \lesssim 10^{10} M_\odot$) is usually expected to decrease with time, the trend is *reversed* (the number of low-mass galaxies is more at lower redshifts $z$) due to the decreasing UV flux with time after $z \sim 2$. We further discuss the implication of our results for the number counts of galaxies and possible evolution of the luminosity function of the galaxies.

*Subject headings:* formation, evolution – galaxies ; diffuse radiation – cosmology

## 1. INTRODUCTION

Formation of galaxies depends sensitively on the dissipative nature of the baryonic gas. While the halos are formed out of the underlying distribution of the dark matter, if there is any, the fate of the resulting galaxy depends much on the behavior of the gas in the halo. Cooling of the gas has long been known to be a decisive factor in determining the characteristic size and mass of the galaxies (Rees & Ostriker 1977; Silk 1977). Any deviation in the history of the universe that alters the state of the gas in halos, therefore, must result in some change in the property and distribution of galaxies.

One motivation to study such effects is the fact that standard model of hierarchical structure formation predicts more low mass galaxies than observed. One then requires a negative feedback mechanism to halt or inhibit the formation of small galaxies (with luminosities less than the characteristic value $L_* \sim 10^{10} h^{-2} L_\odot$), so that most of the mass of the universe is not locked away in such galaxies. Two mechanisms have been put forward to achieve this: heating by supernovae from the first bout of star formation and heating by the UV background radiation. Dekel and Silk (1986) (and recently elaborated upon by Lacey & Silk 1991) have considered the first mechanism in detail. They found that galaxies with velocity dispersion less than $\sim 100$ km s$^{-1}$ are vulnerable to substantial mass loss due to supernovae driven winds. With this limiting value of velocity dispersion, and assuming that dwarf galaxies form out of $1\sigma$ density fluctuations, they derived several scaling relations and compared them with observations.

The possibility that UV photons can inhibit the collapse of small galaxies has been recognized by various authors (e.g. Dekel & Rees 1987). However, attempts have been made only recently to measure, and calculate from theoretical arguments, the intensity of the UV background radiation. While the precise measurements are still eagerly awaited, it seems almost certain that a background radiation with intensity (at the Lyman limit) $J \sim 10^{-22} - 10^{-21}$ erg cm$^{-2}$ s$^{-1}$ sr$^{-1}$ Hz$^{-1}$ permeated the universe at redshifts $5 \gtrsim z \gtrsim 2$ (see, for example, Bechtold *et al.* 1987; Espey 1993). The UV photons are thought to have been produced by quasars and, possibly, young galaxies (Miralda-Escudè & Ostriker 1990; Madau 1992).

Since gas with $T \lesssim 10^{5.5}$ K cools mainly by line cooling of H and He$^+$ atoms, the cooling efficiency is diminished in the presence of the ionizing photons. Efstathiou (1992) calculated the cooling function and, after comparing the cooling time with the Hubble time, argued that galaxies with velocity dispersion $\lesssim 50$ km s$^{-1}$ will not be able to cool and collapse. However, it is not clear as to how much the UV background affects the *mass function of the galaxies*, and the *evolution of the mass function with time* (since the background radiation changes with time). In the hierarchical model of structure formation, density perturbations collapse to form halos at different redshifts depending on the amplitude of fluctuation. Halos with very large amplitude, *i.e.* large density contrast, will be able to self-shield themselves against the UV background and can collapse even if the velocity dispersion is small. Therefore, without detail calculations, one cannot answer the question – to what extent does the UV background inhibit galaxy formation, say, at a certain mass range, if at all?

We believe that such questions are very important, since suppression of galaxy formation via photoionization has been invoked in various contexts. To explain the paradoxical observations of galaxy counts at faint magnitudes (e.g. Cowie, Songaila, & Hu 1991), Babul and Rees (1992) have argued that small mass objects of mass $\sim 10^9$ M$_\odot$ collapse at $z \sim 3$ but do not form stars until $z \sim 1$ due to the inhibiting UV background radiation. After the starburst, galaxies with dispersion velocity less than the Dekel and Silk limit of $v \sim 100$ km s$^{-1}$ are destroyed so that not many of the initial set of galaxies are observed today. The effect of the UV background radiation on the abundance of small mass galaxies, therefore, are important to understand in order to make any model for the faint blue galaxies.

In this paper, we discuss the implications of the UV background in more detail. Our principal goal is to make the estimates and guesses of the previous workers for the effect of UV radiation on galaxy formation more quantitative. We take advantage of the recent measurements of the UV background to calculate the exact changes, some of them hitherto unforeseen, caused by the UV photons in the mass function of galaxies. For simplicity, and because the power spectrum in cold dark matter (CDM) model has been calculated in more detail than in any other model, we will use the CDM spectrum, with $\Omega = 1$. For the Hubble constant, we use $h_{50} = 1$, where $H = 50 h_{50}$ km s$^{-1}$ Mpc$^{-1}$. The structure of the paper is as follows: in §2 we calculate the cooling function, and discuss self-shielding of clouds against UV photons. In the density-temperature plane, we plot the loci of constant amplitude density perturbations and determine which objects can form a neutral core and thus cool. In other words, we determine the threshold $\nu_{th}(M)$ for the amplitude of fluctuation as a function of the UV background intensity. In §3 we use the Press-Schechter and the peak formalism of Bardeen *et al.* (1986) to calculate the mass function of galaxies with and without the UV background. We discuss the implications of our results in §4.

## 2. UV BACKGROUND RADIATION AND EVOLUTION OF GAS CLOUDS

### 2.1 *UV background radiation*

It has been evident since long that the intergalactic gas is highly ionized at high redshift (Gunn & Peterson 1965). Webb *et al.* (1992) have recently claimed to have measured an optical depth of $\tau \sim 0.04$ at $z = 4$. Jenkins & Ostriker (1991) put an upper limit of the optical depth due to neutral hydrogen in the intergalactic medium of $\tau < 0.14$ at $z = 4.2$. While it is possible that the gas could be ionized due to processes other than photoionization, observations of Lyman-$\alpha$ clouds independently suggest the existence of a UV background radiation from the 'proximity effect'. The accurate intensity of the ionizing radiation is yet to be confirmed: the reported flux varies from (in the units of $J_{-21}$, where $J \equiv J_{-21} \times 10^{-21}$ erg cm$^{-2}$ s$^{-1}$ sr$^{-1}$ Hz$^{-1}$) $\sim 1$ ($1.8 < z < 3.5$) (Bajtlik *et al.* 1988) to $\sim 0.1$ ($z \sim 2$) (Dobrzycki & Bechtold 1991; Espey 1993). On the theoretical front, Miralda-Escudè and Ostriker (1990) suggested that quasars and young galaxies could give rise to a flux of $J_{-21} \sim 1$. Madau (1992) calculated the contribution of the quasars and concluded that quasars solely could account for the ionizing flux if $J_{-21} \sim 0.1$ at $z \sim 2$. His calculations suggest that the intensity is almost a constant for $z \gtrsim 2$ and decreases rapidly at lower redshifts. For the present day value of $J_{-21}$, we will use the observations of the sharp edge of HI in spiral discs by Maloney (1992, 1993), who inferred that $J_{-21}(z = 0) = 0.033 \times 10^{\pm 0.7}$.

Below, we will assume a spectral index of the UV radiation $\alpha = 1$ ( i. e., $J_\nu \propto (\frac{\nu}{\nu_H})^{-1}$). Since Black (1981) has calculated the atomic processes for $\alpha = 1$ (which we will use to construct the cooling function), and since, the expected quasar spectrum also has a similar index ($\alpha \sim 0.5$), our assumption makes the calculation simple.

### 2.2 *Cooling rate*

The cooling rate of a plasma at $T \lesssim 10^{5.5}$K with primordial abundance (X=0.76, Y=0.24) depends mostly on the fractional ionization of H and He$^+$. In the absence of any ionizing flux, the cooling is dominated by line emissions from H and He$^+$ atoms. The cooling is therefore highly diminished when UV photons exist (which ionize these atoms) and the cooling is then dominated by recombination. The cooling rates in the presence of ionizing radiation with intensity $J_{-21} = 1, 10$ were calculated by Efstathiou (1992) using the processes tabulated by Black (1981). We have also calculated the cooling rate using the same processes. We find a slight discrepancy, by a factor $\lesssim 1.3$, with that shown in Efstathiou's (1992) paper and realize that such discrepancies abound in the literature (compare, e.g., the cooling rate in Efstathiou 1992 for no ionizing flux with that in Fall &

Rees 1985, both calculated for a primordial gas). We should, therefore, bear in mind the uncertainties when making any argument based on the cooling rate.

The heating rate is also determined by using the processes in Black (1981). The equilibrium temperature of the plasma in the presence of ionizing flux is calculated by balancing the heating and the cooling rate.

### 2.3 Cooling time vs. collapse time

Efstathiou (1992) plotted the limiting curves where cooling time, $t_{cool}$, equals the Hubble time, $t_H$, at various redshifts to show that photoionization can be important as an inhibiting mechanism. We have instead decided to plot curves comparing cooling time and the collapse time which are independent of redshifts (compared to various plots for different redshifts in Fig.2 of Efstathiou 1992).

Fig. 1 is a plot in the particle density ($n$) and temperature ($T$) plane. Cooling time, $t_{cool}$ is defined as (Efstathiou 1992)

$$t_{cool} = E\left(\frac{dE}{dt}\right)^{-1} = \frac{3}{2}\frac{1}{\mu^2(1-Y)^2}\frac{kT}{nL(T,n)}, \tag{1}$$

where $L(T,n)$ (erg cm$^3$ s$^{-1}$) is the cooling function, $\mu$ is the mean molecular weight ($=0.59$ for a totally ionized gas) and $n_H = \mu(1-Y)n$ is the number density of hydrogen atoms. The time scale of collapse of an object, $t_{ff}$ is given by (Dekel & Silk 1986),

$$t_{ff} = (6\pi G\rho)^{-1/2} = 8.9 \times 10^{14}\ n^{-1/2}\ \text{s}, \tag{2}$$

where we have assumed that the gas mass comprises only 10% of the total mass.

The $t_{cool} = t_{ff}$ curves for $J_{-21} = 0, 0.1, 1$ are shown in Fig.1 (when comparing with Fig.5 of Dekel and Silk (1986), note that their $n$ is actually $n_H$ in our definition).

We have plotted the curves for constant amplitude of the density fluctuations in CDM model. The curve marked '$1\sigma$' denotes the perturbations for which $\delta M/M = 1$ at a comoving radius of 16 $h_{50}^{-1}$ Mpc. We have used the fact that, if $\delta_0$ is defined as the linear density fluctuation extrapolated to the present time, then $\delta_0 = \nu\Delta(M)$, where $\Delta(M)$ is the variance (an analytical approximation is given in White & Frenk 1991). Spherical top-hat model gives,

$$T = \frac{\mu m_p}{3k}\left(\frac{4\pi}{3}\right)^{1/3}GM^{2/3}\rho^{1/3} \tag{3a}$$

$$\rho = 18\pi^2\rho_0\left(\frac{\nu}{1.686}\right)^3\Delta^3(M), \tag{3b}$$

where $m_p$ is the proton mass and $\rho_0$ is the present mean density of universe with $h_{50} = 1$ and $\Omega = 1$. These relations allow us to draw the constant amplitude curves in the $n - T$ diagram. One can scale the curves for other values of the biasing parameter, $b$ (we have assumed $b = 1$ for Fig.1), using $\delta_0 \propto \nu/b$.

The corresponding curve (separated by an arrow) denotes the gas clouds after a contraction by a factor of $F^{-1/3}$, or equivalently the increase of density by a factor of $F$, where $F$ is the ratio of gas mass to the total mass (on the assumption of uniform density). The gas densities on this curve are comparable to the halo densities so that star formation can ensue (White & Rees 1978; Dekel & Silk 1986). We shall next require that the gas clouds, after a collapse to this stage, be self-shielded against UV photon to ensue cooling and star formation.

The density for the above curve is calculated for a uniformly distributed gas in CDM halos, with the fraction of gas mass being 0.1. The short-dashed diagonal lines denote objects of a given mass (again with a gas mass fraction of 0.1).

The equilibrium temperature $T_{eq}$ is then plotted as a function of the density for $J_{-21} = 0.1, 1$. It is easily seen that for the objects that can cool faster than their collapse time, $T_{eq}$ ranges between $10^4 - 10^5$ K.

### 2.4 Self-shielding

Large objects can shield themselves against the UV background radiation. For perturbations with large density, the optical depth to the UV photons can reach unity at a certain radius and the gas inside may be shielded against any UV photons and can, therefore, cool fast. There exists a critical radius for which the rate of photoionization ($\propto r^2$, the surface area) balances the rate of recombination ($\propto r^3$, the volume). The critical radius, $r_c$, is given by the balancing equation (assuming a spherical and uniform gas cloud),

$$\frac{4\pi}{3} r_c^3 n_e n_p \alpha(T) \approx (4\pi r_c^2) \, 1.5 \times 10^5 J_{-21} \,, \tag{4}$$

where $\alpha(T)$ is the recombination rate for hydrogen atoms, and $n_e, n_p$ are the electron and proton densities. We use $T_{eq}$ in the above equation, assuming that objects with larger virial temperatures will cool down ($t_{cool}$ is much shorter than the Hubble time for objects inside the region $t_{cool} < t_{ff}$) and those with lower virial temperatures will be heated up to $T_{eq}$ quickly.

The sphere with the critical radius, $r_c$, is essentially an 'inverted Strömgren sphere'. Objects bigger than this will have a core where no UV photons can reach, because the surrounding mantle of ionized gas balances the photoionization with recombination. (The $r_c$ calculated here refers only to hydrogen; however, there would be a thin shell where both H and $He^+$ will be ionized, surrounding a shell where H is ionized but $He^+$ will be neutral. We neglect such details, since recombination due to H is dominant.)

The virial temperature of the object with the critical radius is easily calculated (total mass being 10 times the gas mass), when they are assumed to be self-gravitating. We plot the curves for these critical objects in Fig.1 for $J_{-21} = 0.1, 1$ (the dotted-and- dashed lines) and will refer to them as the 'self-shielding curves'.

One can calculate the core mass for objects lying above the self-shielding curves, by calculating the thickness of the shell, $\Delta r$, where recombination and photoionization are balanced. One can easily show that

$$\Delta r = R - \left[R^2 \left(R - \frac{3(4\pi)\, 1.5 \times 10^5 J_{-21}}{n_e\, n_p\, \alpha(T)}\right)\right]^{1/3}, \qquad (5)$$

where $R$ is the radius of the gas cloud, and $n_e, n_p, \alpha$ are expressed in cgs units. Fig. 2 compares the core mass (after a dissipative contraction by a factor $F^{-1/3}$) with the total mass for different values of $J_{-21}$ and $\nu$. The points where the core masses plunge to zero are essentially the points of intersection of constant $\sigma$ curves after a contraction with the self-shielding curves in Fig.1.

Note that in deriving eq.(5) we have only considered photoionization. Collisional ionization would be important for clouds with $T_{eq} > 10^5\,\mathrm{K}$ and would lower the self-shielding curves at the high temperature region of the $n - T$ diagram. We will neglect this effect in the following and only note here that it would make our results more conservative.

### 2.5 Implications of the $n - T$ diagram

The cooling diagram Fig.1 shows the features that have been discussed qualitatively by previous works, *viz.*, the suppression of the dwarf galaxies. If the dwarf galaxies originate from typical $1\sigma$ fluctuations, as is commonly believed (Dekel & Silk 1986; Babul & Rees 1992), then their formation will be inhibited by the UV photons if $J_{-21} \sim 10$. Figure 1 shows that for $J_{-21} \sim 1$, objects with $\nu \sim 0.5$ will not be able to self-shield and cool, after collapsing by a factor of $F^{-1/3}$. With the decrease of the UV flux after $z \sim 2$, the self-shielding curves will be lowered in Fig.1, and objects that were earlier above the curves will now be able to form neutral cores and harbor star formation.

¿From Figs. 1 and 2, it is quite obvious that for a given amplitude of the density fluctuation, $\nu$, objects with larger masses (and therefore with smaller densities) are more vulnerable to the UV background radiation. These objects remain ionized and the heating due to the UV photons keep the temperature at $\gtrsim T_{eq}$ and cannot cool. In other words, one can say that for a given mass $M$, the value of $\nu$ must be bigger than some lower limit, $\nu_{th}(M)$. This threshold value of $\nu$ is shown in Fig.4 as a function of the total mass and $J_{-21}$ (see the next section for details). Armed with the knowledge of $\nu_{th}(M)$ we can now calculate the effects of the UV background on the mass function of galaxies in the next section.

## 3. GALAXY MASS FUNCTION IN THE UV BACKGROUND

In the presence of photoionizing radiation, the number of galaxies which can cool before a given epoch is reduced. Since the effect depends on mass, as we have seen in the previous section, the shape of mass function of galaxies will inevitably be changed by the UV background. Blanchard *et al.*(1992) have also discussed, using the different model, the effects of gas heating on the galaxy mass function. In this section, we describe the basic formalism to derive the mass function and discuss the results. We will not touch upon the luminosity function of galaxies, which can be deduced only by introducing models of star formation (e.g. Lacey & Silk 1991; White & Frenk 1991, Lacey *et al.* 1993). Since the exact theory of star formation still eludes us, we will only derive the mass function and attempt some model-independent conclusions for the luminosity function.

### 3.1 *Basic equations*

We calculate the mass function using Press-Schechter formalism (Press & Schechter 1974, hereafter PS) and the peaks formalism (Bardeen *et al.* 1986). In the PS formalism, mass function of galaxies at a fixed time is determined by earlier history of hierarchical mergings —i.e. the number of galaxies is reduced by their mergers into larger collapsed systems. In the peaks theory, on the contrary, the peaks which will eventually be within the larger peaks are counted as independent objects —i.e. no merging is assumed.

PS formalism gives an analytical description for the number of galaxies at a given epoch which have not been incorporated into the collapsed larger system. It applies to the evolution of dissipationless dark halos, which are subject to hierarchical merging as seen in cosmological N-body simulations (e.g. Frenk *et al.* 1988). It is however not obvious as to whether baryonic cores, once formed, merge together as rapidly as the dark halos do. The small cross sections for core-core interaction would reduce the rate of merging, so that one-to-one correspondence between dark halos and baryonic cores may not be established (see, e.g. Carlberg & Couchman 1989; Carlberg 1993). This is the reason why we also use the peaks formalism in which no merging is assumed to estimate the mass function. The actual number of galaxies at a given mass probably lies between the values predicted by these two formalisms.

In CDM model, it is assumed that galaxies arise from the linear density fluctuation $\delta \equiv \delta\rho/\bar{\rho}$ with Gaussian statistics, where $\bar{\rho}$ is the mean density of Universe. The idea is that if the mean linear overdensity extrapolated to $z = 0$ is larger than some critical $\delta_c$, then the (spherical) perturbation will have eventually collapsed and virialized. We adopt $\delta_c = 1.69$ as given by the spherical "top-hat" collapse model.

In the standard PS formalism, the number of objects per comoving volume per unit mass is usually written as,

$$n(M,z)dM = \left(\frac{2}{\pi}\right)^{1/2} \rho_0 \frac{\partial \ln \Delta}{\partial \ln M} \nu \exp(-\nu^2/2) \frac{dM}{M^2}, \qquad (6)$$

where $\rho_0$ is the present mean density of Universe; $\nu$ is the amplitude of density perturbation for the onset of collapse in the units of rms density fluctuation, and is defined as

$$\nu = \frac{\delta_c(1+z_c)}{\Delta(M)}. \qquad (7)$$

where $z_c$ is the epoch of collapse. $\Delta(M)$ is the rms linear overdensity for the mass scale $M$ at the present time. Equation (6) applies to galaxies which have not been incorporated into larger collapsed system at a given $z_c$. Here we note that one must be careful about this standard form: eq.(6) is derived on the assumption of constant critical $\delta_c$. However the adoption of the cooling and self-shielding criterions as we will consider below introduces the mass (or equivalently, the filter radius)-dependent critical overdensity $\delta_c(M)$, and the simple incorporation of this effect into eq.(6) is beyond its range of validity. This invalidity of the standard PS formula in general cases of $\delta_c(M)$ is related to the factor of 2 fudge to derive eq.(6): half the mass which is uncounted in PS, i.e. underdense regions with $\delta < \delta_c$, can have $\delta > \delta_c$ for some larger filter in general.

The explicit consideration of this effect in the PS formalism was given by Peacock & Heavens (1990) and Bond *et al.* (1991), and we will adopt the former recipe, which can apply to general cases of $\delta_c(M)$. Denoting by $p_G(\delta > \delta_c)$ the usual Gaussian probability that a point has $\delta > \delta_c$ for a given filter radius $R_f$, Peacock & Heavens showed that the probability $p(R_f)$ that the point is ultimately incorporated into a larger $R_f$ system is given as,

$$p(R_f) = p_G + (1-p_G)(1-p_s), \quad (8)$$

where $p_s(R_f)$ is the survival probability that the point is always below $\delta_c$ for larger $R_f$, given in eq.(11) and (14) by Peacock & Heavens (1990). The differentiation of this by mass defines the mass function per unit volume

$$Mn(M,z)dM = \rho_0 |\frac{dp}{dM}|. \quad (9)$$

On the other hand, in the peaks formalism, the differential number density of peaks for a given $R_f$ is given as

$$\frac{dn_{pk}}{d\nu} = \frac{1}{(2\pi)^2 R_*^3} \exp(-\nu^2/2) G(\gamma, \gamma\nu). \quad (10)$$

where $R_*$ and $\gamma$ are defined in terms of a set of spectral moments $\sigma_i$, i.e. $R_* = \sqrt{3}\sigma_1/\sigma_2$, and $\gamma = \sigma_1^2/\sigma_0\sigma_2$, respectively, and $G$ is defined in Bardeen *et al.* (1986). For the differential number density in the peak size $R_f$, we adopt the approximate approach given by Lacey & Silk (1991):

$$\frac{d^2 n_{pk}}{d\nu dR_f} = -\frac{1}{(2\pi)^2} \frac{3}{R_f^4} \left(\frac{R_f}{R_*}\right)^3 \frac{dlnR_*}{dlnR_f} \exp(-\nu^2/2) G(\gamma, \gamma\nu), \quad (11)$$

and integrate this over the peak height $\nu$ to obtain the mass function.

We use the power spectrum derived by Davis *et al.* (1985). It is important to notice that if we adopt the top-hat filter in these two formulations, which has a sharp edge in $r$-space but is oscillatory in $k$-space, the integrals for spectral moments do not necessarily converge for all possible slopes of the fluctuation spectrum. The Gaussian filter avoids this undesirable feature, and thus we adopt it, using the same power spectrum. In order to relate the ingredients of top-hat collapse model, such as $\delta_c = 1.69$ for collapse, to the Gaussian filtering, we have normalized the Gaussian-smoothed $\Delta(M)$ to be unity at the corresponding top-hat lengthscale $R$ of the usually adopted value $16 h_{50}^{-1}$ Mpc. The mass of a Gaussian-filtered peak is estimated by $M = (2\pi)^{3/2} R_f^3 \rho_0$, so that the relation between $R$ and $R_f$ is given by $R = (3/4\pi)^{1/3}(2\pi)^{1/2} R_f$. In Fig.3, we have plotted the two $\Delta$'s, as functions of the top-hat lengthscale. The difference between the two cases is small. Of course, assigning a proper volume to the Gaussian filter function in this way is not rigorous and is still unresolved (Peacock & Heavens 1990). However the slight difference in the results by adopting the different filters does not cause any essential change in the main conclusions drawn in the later sections.

### 3.2 The effects of cooling and UV background

The mass functions with the threshold value for $\nu$ given in eq.(7) apply only to objects which can collapse and virialize before a given redshift. In order to form stars, however, the hot interstellar gas with virial temperature, $T \sim 10^{5-6} K$, must cool via radiative processes. Furthermore, in the presence of UV background, the gas must be dense enough to form a neutral core.

These two requirements on the virialized objects can be expressed in terms of the mass-dependent thresholds, $\nu^{cool}(M)$, below which the objects cannot cool before a given $z$ in the absence of UV background, and $\nu^{UV}(M)$, below which they cannot form self-shielded cores against UV photons. We define $\nu^{cool}(M)$ for the given total mass $M$, such that the cooling time given in eq.(1) (without the effect of photoionization) satisfies the following condition,

$$t_{cool} \lesssim \frac{2}{3H_0}[(1+z)^{-3/2} - (1+z_c)^{-3/2}], \tag{12}$$

where $z_c$ is the redshift of collapse. At the high mass end for $M \gtrsim 10^{10} M_\odot$, line cooling by metals produced in galaxies formed earlier is important. To include the effect of metal cooling, we have adopted the simplification, as in Peacock and Heavens (1990), that for a cooling function dominated by recombination below $T \sim 10^8$ K, the collapsed redshift can be scaled as

$$(1 + z_c) = (1 + z)(1 + M/M_{cool})^{2/3}, \tag{13}$$

where

$$M_{cool} \sim 1.68 \times 10^{13} f_m \ M_\odot. \tag{14}$$

The parameter $f_m$ denotes the metal content: $f_m = 1$ for solar abundances and $f_m \sim 0.03$ for no metals. We have assumed $f_m = 1$ for metals and 10% for the gas fraction. $M_{cool}$ then is the typical mass scale below which cooling is important. Substituting the above expression for $(1 + z_c)$ in eq.(7), the threshold for the density fluctuation in eq.(7) will have an extra factor for $\delta_c$, i.e. $\delta_c \to \delta_c(1 + M/M_{cool})^{2/3}$, indicating the increase of the threshold due to the cooling condition. At the low mass end for $M \lesssim 10^{10} M_\odot$, since the effect of metalicity is trivial, we have determined $\nu^{cool}(M)$ based upon the zero-metal cooling rate. As we will show below, the condition to form a shielded core represented as dash-dotted curves in Fig.1 is more stringent than the cooling condition. Our use of two different cooling functions in two different mass ranges does not affect the final result for the mass function of galaxies.

The effect of UV background on the threshold, $\nu^{UV}(M)$ was obtained in §2. Fig.2 shows that the shielded core mass (including both dark halos and baryons) is almost equal to the total mass for $\nu(M) \geq \nu^{UV}(M)$, and is zero for lower values of $\nu(M)$.

The new threshold value for density fluctuations which can self-shield and cool, is then given as

$$\nu_{th}(M) = max(\nu, \nu^{cool}, \nu^{UV}). \tag{15}$$

Fig.4 shows the threshold values for collapse and cooling of objects by the present epoch $z = 0$, $\nu^{UV}$ for self-shielding in the UV flux $J_{-21} = 0.1$, and the adopted $\nu_{th}$ given in eq.(15), as a function of core mass $M_b$. As discussed in §2, $\nu^{UV}$ is calculated on the assumption that the baryonic gas with the fraction, $F = 0.1$, in mass, collapses by a factor $F^{-1/3}$ within a uniformly distributed dark halo. At both high-mass $M_b \gtrsim 10^{12} M_\odot$ and low-mass $M_b \lesssim 10^7 M_\odot$ ends, $\nu_{th}$ is very large because of the cooling constraint. It is also evident that the condition to form a shielded core against UV photons essentially determines the $\nu_{th}$ in the mass range $10^8 \lesssim M_b \lesssim 10^{12} M_\odot$, i.e. from dwarf to bright galaxies. This is clearly seen in Fig.4b where the dependence of $\nu_{th}$ on the UV flux $J_{-21}$ is shown at a fixed mass scale $M_b = 10^{11} M_\odot$. It shows that $\nu_{th}$ increases with $J_{-21}$.

Having obtained $\nu_{th}(M)$ for several constraints, the number density of objects as a function of core mass $M_b$ can be calculated based upon eq.(9) and eq.(11) for the generalized PS and peaks formalisms, respectively.

### 3.3 Mass function at z=0

In Fig.5, we show the mass functions of baryon masses $M_b$ at $z = 0$ under the above conditions, based upon the PS and peaks formalisms. The UV flux, $J_{-21}$, is assumed to be 0.1. The mass functions for collapsed objects or peaks have the power-law tail $dn/dM_b \propto M_b^{-2}$ on small mass scales. The exponential cutoffs at high-mass ends exist simply because the objects with $\delta < \delta_c$ for all larger radii cannot collapse by the present time. The mass function in the peaks formalism has a slightly steeper slope at low-mass ends than that in the PS formalism because of counting the small-scale peaks which are actually within larger systems. The amplitude of mass function at low-mass ends is obviously related to the efficiency of merging.

The mass function with cooling condition denotes the objects which have cooled by the present time. In both formalisms, the numbers of high mass objects are drastically cut off due to ineffective cooling at these mass scales. In the PS formalism, the number density at intermediate mass $M_b \sim 10^{12} M_\odot$ is increased, because the ineffective cooling of more massive objects inhibits the merging of objects at this mass scale and their survival increases their number density (see Peacock & Heavens 1990). For low-mass scales $M_b \lesssim 10^7 M_\odot$, where the virial temperature is below $T \sim 10^4$K and cooling is negligible, there appears a cutoff in both formalisms. The ineffective cooling in these mass scales leads to the high threshold value $\nu$ (Fig.4a) for density fluctuation, resulting in the suppression of low-mass galaxy formation. According to J.Peacock (private communication), this decrease of the probability in the PS formalism that low-mass objects can cool is compensated by the high probability for larger objects, and this is the origin of the bump just above this cutoff of the mass function — as also happens just below the high-mass cutoff. Note that even if the present recipe fails to derive mass function at low-mass ends, such very low-mass objects are observationally unimportant due to the selection effects (Yoshii 1993).

Under the cooling condition, there is a slight dip in the PS mass function at $M_b \sim 10^{8.5} M_\odot$. This is artificially produced in our method, because our $\nu_{th}$ given in eq.(15) and Fig.4a has a discontinuous change from the cooling-condition dominant (low-mass) scales to the collapse-condition dominant scales (at $M_b \sim 10^{8.5} M_\odot$). Probably, nature would avoid such a discontinuity in $\nu$, e.g. due to the effects of non-uniform and non-spherical density distribution of gas, and smooth out the mass function. However such a feature in the mass function does not affect the main conclusions below.

Imposing the more stringent condition of forming self-shielded cores in the presence of UV background leads to further decrease in the mass function. Table 1 and 2 give the number density as a function of the UV flux for a few mass scales. It is evident that at the high-mass end, the number of objects which can form a core is smaller than that of objects which can cool by the present time. This is due to the efficient photoionization of the low-density gas in the massive objects (see Fig.4). For masses $M_b \gtrsim 10^{13} M_\odot$, however, the cooling condition becomes more important, giving rise to a sharp cut off. At low-mass end, the peaks formalism predicts a decrease in the number density, whereas it does not change much in the PS formalism. As a result, the slope of the mass function in PS formalism becomes steep due to the UV background, while it is slightly steepened in the peaks formalism. An artificial irregularity in the PS mass function, as happened in the case without UV, also appears at $M_b \sim 10^8 M_\odot$ in Fig.5a, where our $\nu_{th}$ given in eq.(15) and Fig.4a has a discontinuous change from the cooling-condition dominant (low-mass) scales to the shielding-condition dominant scales.

To estimate quantitatively the change of slope, we assume a power-law form, $dn/dlnM_b \propto M_b^{\alpha+1}$, for the mass function at low-mass ends $10^8 M_\odot \lesssim M_b \lesssim 10^{10} M_\odot$, and fit the derived mass function. $\alpha$ is a constant index denoting the slope of mass function at low-mass ends. In table 3, we show the values of $\alpha$ as a function of $J_{-21}$. $\alpha$ is apparently a decreasing function with $J_{-21}$. It is interesting to note that the resultant shape and slope of tions in the UV background are similar for both formalisms, albeit with higher amplitude in PS-formalism.

### 3.4 Evolution of the mass function

In order to calculate the evolution of mass function in the presence of the UV background, we need to express the possible evolution of the UV flux, $J_{-21}$, as a function of $z$. As we mentioned in §2, several observations and calculations suggest that $J_{-21}$ is almost constant before $z \sim 2$, and rapidly declines at lower redshifts. Since the knowledge of $J_{-21}$ at intermediate redshifts is still uncertain, we have adopted a simple functional form for $J_{-21}(z)$:

$$J_{-21}(z) = \begin{cases} J_{-21}(z=2) = const. & \text{for } z \geq 2 \\ J_{-21}(z=2) \left(\frac{1+z}{3}\right)^\alpha & \text{for } z < 2. \end{cases} \quad (16)$$

We adopt 3 characteristic cases for the evolution of $J_{-21}$, defined by $(J_{-21}(z=2), J_{-21}(z=0)) = (1.0,0.1)$, $(3.0,0.1)$, and $(3.0,0.05)$. The parameters $\alpha$'s in eq.(16) are then equal to 2.10, 3.10, and 3.73 respectively.

In Figs.6 and 7, we show the redshift dependence of mass functions in the range $0 \leq z \leq 3$ for PS and peaks formalisms, respectively. The collapse factor of baryonic gas is again $F^{-1/3}$ for the cases with UV background. The major difference in the mass functions between the two formalisms is that when there are no constraints due to cooling and UV background, the amplitude of mass function in PS formalism is a decreasing function with time (Fig.6a), whereas the number of collapsed peaks is always increasing (Fig.7a). This difference is due to the explicit consideration of merging in the former case. The characteristic mass $M^*$, (where the power-law mass function turns to an exponential fall-off at high-mass end), increases in both formalisms, since larger systems collapse at later times.

However, the effect of photoionization tends to smear out such a difference (Figs.6b, 6c, 6d, 7b, 7c and 7d). The evolution of mass function is greatly influenced by UV, especially in the mass range $10^8 \lesssim M_b \lesssim 10^{10} M_\odot$: the amplitude of mass function at a given mass scale tends to increase with time after $z = 2$, even in PS formalism. The slope is slightly steeper at higher $z$, because the lower-density gas in more massive objects is more vulnerable to photoionization. Before $z = 2$, the shape of mass function does not change with time owing to the constancy of the UV flux at these epochs. The rate of increase in the amplitude is dictated by the rate of decrease in UV, i.e. the evolution of the mass function in time is greater for larger $J_{-21}$ at $z = 2$ (for the same value $J_{-21}(z = 0)$).

The effects of UV upon the mass function obtained here do not depend upon an imposed bias parameter $b$. If we decrease the value of $b$ from 2 to 1, the threshold value of density fluctuation, $\nu_{th}$, is decreased as shown in Fig.4(b), and the amplitude of mass function is accordingly increased. However, the steepened slope and the decreased amplitude of the mass function due to UV background are unchanged by this parameter.

## 4. DISCUSSION

### 4.1 *Galaxy luminosity function*

To compare our results for the mass function with observations, one needs to derive the luminosity function from the mass function. The mass-to-luminosity ratio $M/L$, however, is generally not a constant in mass and time and detailed modeling of star formation and gas dynamics such as gas loss from galaxies are required to derive the actual luminosity function (e.g. Lacey & Silk 1991; White & Frenk 1991). Such a derivation of the luminosity function is therefore very model dependent.

The possible model-independent effects of the UV background upon the luminosity function are: (i) the number density of galaxies decreases due to UV, if one adopts the same star-formation law for both models with and without UV; (ii) the slope of the function would change at a mass scale, say $M_s$, where the UV-dominant regime for $M_b \lesssim M_s$ turns to the cooling-dominant one for $M_b \gtrsim M_s$ ($M_s \sim 10^{12.5} M_\odot$ in Fig.5). The mass-dependent $M/L$ would not change this property, because it is unlikely that this ratio varies by an order of magnitude in the narrow range of mass scales near $M_s$, as is the case for the mass function (note the log-scale ordinate in Fig.5). We will further discuss the effect (i) later in conjunction with the galaxy number counts.

### 4.2 *A biasing mechanism*

The necessity of some physical biasing mechanism in galaxy formation has been argued by various authors in trying to reconcile the CDM model with large-scale galaxy distribution and velocity distribution. Rees (1984) had argued that photoionization should make it 'difficult' for galaxies, specially those with $\nu \lesssim 1$, to collapse. We have been able to quantify the 'difficulty' in terms of the threshold amplitude of fluctuations, $\nu_{th}(M)$. As Fig.4b shows, the threshold amplitude is indeed increasing with $J_{-21}$, which results in the suppression of the number density of baryonic cores (Table 1). We therefore have a natural biasing of the galaxies relative to the dark halos. For instance, according to Fig.4b at the fixed mass scale $M_b = 10^{11} M_\odot$, the value of $\nu_{th}(z=0)$ for $J_{-21} = 0$ with a bias parameter $b = 2$ is comparable to that for $J_{-21} = 0.05$ without artificial bias, $b = 1$. Furthermore as is shown in Fig.4a, the threshold amplitude tends to increase more at higher masses due to the UV background, simply because the low density gas at high mass scale is easily photoionized.

Therefore, the presence of the UV flux in the background provides a biasing mechanism for the formation of galaxies: the effect is more at brighter galaxies, enhancing the clustering. This results in the preferential formation of bright galaxies in clusters and superclusters. It is also possible that the spatial distribution of quasars, which are the main contributors to the UV flux, introduces the modulations in the galaxy distribution (Efstathiou 1992). Bahcall and Chokshi (1991) suggested that quasars are preferentially located in small groups of galaxies ($N = 10 \sim 30$), while according to Mo and Fang (1993), quasars trace the same large-scale structure of galaxies and cluster of galaxies. If quasars are located in high-density environment of galaxies, the UV flux there would be stronger than that in the field. This will thus result in steeper slope of the mass function in the richer environments due to stronger UV flux (see Table 3). The observed correlation between the slope of luminosity function and the richness of group (e.g. Ferguson 1992 for a review) may be explained by such an inhomogeneity of the UV flux.

It is true that our calculations for uniform spheres may not be realistic in the sense that one expects the density profile in the halos to be more like isothermal. The collapsing factor would be $F^{-1}$ in that case, instead of $F^{-1/3}$ that we used. It is however not clear, without a detail computation that includes the density profile, how the density profile affects the results.

### 4.3 Galaxy number counts

The presence of the UV background has further implications in the interpretation of galaxy number counts. The number count at an apparent magnitude $m_\lambda$ at $z$ (where the subscript $\lambda$ denotes the wavelength dependency), with an absolute magnitude $M_\lambda$, is given by the product of the luminosity function $\phi(M_\lambda, z)$ and the volume element $dV/dz$ of the comoving space:

$$N(m_\lambda, z) = \frac{\omega}{4\pi} \phi(M_\lambda, z) \frac{dV}{dz}, \qquad (17)$$

where $\omega$ is the angular area in units of steradians over which galaxies are counted. Integrating over $z$, we obtain the differential number count of galaxies $N(m_\lambda)$ as a function of $m_\lambda$. $N(m_\lambda)$ is strongly dependent on the evolution of galaxies through $M_\lambda(m_\lambda)$ relation and the evolving $\phi$, and on the cosmological parameters through volume effects (e.g Yoshii & Takahara 1988; Rocca-Volmerange & Guiderdoni 1990). In order to explain the large number of blue faint galaxies revealed by the observation (Tyson 1988), several attempts have been made, which include: a merging-driven evolution with $\Omega = 1$ based upon the Schechter-form luminosity function (Rocca-Volmerange & Guiderdoni 1990); the effect of a large cosmological volume element (low $\Omega$ and non-zero cosmological constant $\Lambda$: Fukugita *et al.* 1990); the bursts of star formation at a recent epoch (Lacey & Silk 1991; Babul & Rees 1992). The interpretation of galaxy number counts is thus highly dependent upon luminosity and number evolutions as well as cosmological parameters.

The mass function based upon the hierarchical clustering of CDM predicts the steep mass spectrum $dn/dM_b \propto M_b^{-2}$ on small mass scales (Table 3; Lacey *et al.*1993), in contrast with the observed slopes in the field, $\alpha \sim -0.95$ (Loveday *et al.*1992) and in the cluster of galaxies, $\alpha \sim -1.3$ (Sandage *et al.*1985). This suggests the excess of low-mass galaxies to explain the faint galaxy count, but contradicts with the present galaxy luminosity function, even considering supernova-driven mass-loss processes (Lacey *et al.*1993).

In the presence of the UV background, the present work suggests that the number density of objects which can collapse and cool by a given epoch is *smaller* than that without UV due to the photoionization effect. Although the accurate incorporation of star formation and gas ejection via supernova-driven galactic wind, which make the slope of luminosity function shallower, is beyond the scope of this paper, the effect of the UV background may decrease the amplitude of luminosity function by reducing the number of the pre-galactic objects. Furthermore, because of the decreasing UV flux with time for $0 \lesssim z \lesssim 2$, the time-evolution of the mass function is *not constant nor decreasing with time as is usually assumed* in the models of number counts (Yoshii & Takahara 1988; Rocca-Volmerange & Guiderdoni 1990). The effect of photoionization is thus suppressing the number of galaxies, preferentially at earlier epochs (Efstathiou 1992).

Although we have investigated only the case of $\Omega = 1$ universe, these effects of the UV background on the mass function are essentially the same for other cosmological parameters. It was proposed, using the Schechter luminosity function $dn/dL \propto L^{-1}$ on low luminosity scales $L$, that the strong number evolution of galaxies via merging, i.e. larger number of building blocks (premergers) at higher redshifts, can save the $\Omega = 1$ universe (Rocca-Volmerange & Guiderdoni 1990). However, our result of increasing mass function amplitude with time by the UV background implies a *larger* discrepancy between observed and theoretically predicted number counts: the number of premergers is smaller at earlier epochs. This *opposite* sense of time-evolution in mass function may not support the assumption of the strong decrease in the number of galaxies adopted in the number counts model. Note here that the presence of a bump in the mass function at low-mass scale $M_b \sim 10^7 M_\odot$ in the PS formalism is not important in this discussion because such low-mass galaxies at high $z$ would not be detected due to the selection effects (Yoshii 1993). Therefore, it is likely that the high density $\Omega = 1$ universe does not reconcile with the large number of faint counts.

We note from fig. 5 that the decrease in the number density of the galaxies is small for reasonable values of $J_{-21}$. This means that invoking the decrease in $J_{-21}$ after redshift of $z \sim 2$ to explain the emergence of a number of small mass galaxies, as was done by Babul and Rees (1992) to explain the galaxy counts at faint magntitudes, can be fraught with difficulties. Our results do not support the idea that only the decrease in the UV background radiation after $z \sim 2$ could trigger an epoch of starbursts at $z \sim 1$ in dwarf galaxies, as is required in the mechanism of Babul and Rees (1992).

## 5. CONCLUSIONS

We have investigated the effects of the UV background radiation at redshifts $z = 0 \sim 3$ on the mass function of galaxies and its evolution in CDM model. Imposing the condition that, in order to cool for star formation, objects should be able to self-shield against photoionization, we have determined the threshold amplitude of density fluctuations. We have used the Press-Schechter and peaks formalisms to calculate the abundances of objects and evolution of the mass function in time.

We found that the number density of objects is reduced in the presence of the UV photons. For example, at $z = 0$, and for $J_{-21}(z = 0) = 0.1$, the number density of $M_b \sim 10^{11}$ M$_\odot$ galaxies is less by a factor of $\sim 2$ compared to the case of no UV radiation.

Our results, however, do not support the claim that an epoch of starbursts at $z \sim 1$ is triggered by the decrease in the UV background radiation after $z \sim 2$, as has been argued to explain faint blue galaxies. We also found that the number density of low-mass objects *increases* in time, in both Press-Schechter and peaks formalism, due to the evolution of the UV flux. This means that the difference between the cases with and without the UV background is *larger* at higher redshifts ($0 < z < 2$). This result is contrary to the usual view voiced in the literature, in the context of merging driven models of galaxy evolution, that number density of small mass objects decreases in time.

We have discussed the implications of our result with respect to galaxy number counts and the luminosity function of the galaxies. We found the effect of the UV background radiation to be large, and it seems important to include its effect in any realistic model of galaxy evolution.

We are indebted to Drs. Yuzuru Yoshii, John Peacock, Alan Heavens, and Peter Biermann for their useful comments and discussions. We are also grateful to the referee for his excellent and constructive suggestions for the improvement of this paper. MC thanks the Alexander von Humboldt Foundation for support and BBN thanks the Max Planck Society for fellowship.

TABLE 1

Changes in the number density (Mpc$^{-3}$) (PS formalism)

| $M_b(M_\odot)$ | $J_{-21} = 0.0$ | 0.1 | 0.5 | 1.0 |
|---|---|---|---|---|
| $10^9$........ | $2.03 \cdot 10^{-1}$ | $2.26 \cdot 10^{-1}$ | $2.29 \cdot 10^{-1}$ | $2.17 \cdot 10^{-1}$ |
| $10^{10}$........ | $3.11 \cdot 10^{-2}$ | $2.81 \cdot 10^{-2}$ | $2.58 \cdot 10^{-2}$ | $2.26 \cdot 10^{-2}$ |
| $10^{11}$........ | $4.56 \cdot 10^{-3}$ | $2.99 \cdot 10^{-3}$ | $2.22 \cdot 10^{-3}$ | $1.80 \cdot 10^{-3}$ |
| $10^{12}$........ | $8.29 \cdot 10^{-4}$ | $2.87 \cdot 10^{-4}$ | $1.35 \cdot 10^{-4}$ | $7.71 \cdot 10^{-5}$ |

TABLE 2

Changes in the number density (Mpc$^{-3}$) (Peaks formalism)

| $M_b(M_\odot)$ | $J_{-21} = 0.0$ | 0.1 | 0.5 | 1.0 |
|---|---|---|---|---|
| $10^9$........ | $2.94 \cdot 10^{-1}$ | $2.00 \cdot 10^{-1}$ | $1.47 \cdot 10^{-1}$ | $1.21 \cdot 10^{-1}$ |
| $10^{10}$........ | $2.71 \cdot 10^{-2}$ | $1.68 \cdot 10^{-2}$ | $1.09 \cdot 10^{-2}$ | $8.84 \cdot 10^{-3}$ |
| $10^{11}$........ | $2.14 \cdot 10^{-3}$ | $1.24 \cdot 10^{-3}$ | $7.12 \cdot 10^{-4}$ | $5.50 \cdot 10^{-4}$ |
| $10^{12}$........ | $1.01 \cdot 10^{-4}$ | $8.12 \cdot 10^{-5}$ | $3.60 \cdot 10^{-5}$ | $2.24 \cdot 10^{-5}$ |

TABLE 3

Changes in the slope of mass function at low-mass scales $10^8 \leq M_b \leq 10^{10} M_\odot$

| Formalism | $J_{-21} = 0.0$ | 0.1 | 0.5 | 1.0 |
|---|---|---|---|---|
| PS.......... | $-1.89$ | $-1.93$ | $-1.96$ | $-1.99$ |
| peaks....... | $-2.02$ | $-2.07$ | $-2.10$ | $-2.12$ |

**Figure Captions**

**Figure 1.** Gas number density versus virial temperature or virial velocity. *Thick solid curves*: the cooling curves where $t_{ff} = t_{cool}$, for $J_{-21} = 0, 0.1, 1$. *Dotted curves*: the equilibrium temperature $T_{eq}$ where cooling rate is equal to heating rate due to photoionization, for $J_{-21} = 0.1$ and 1. *Dash-dotted curves*: the self-shielding curves against UV photons for $J_{-21} = 0.1$ and 1. *Solid curves*: the loci of density perturbation for $1\sigma$ in the virial equilibrium normalized to unity at a comoving radius $16h_{50}^{-1}$Mpc, with a bias parameter $b = 1$. The corresponding parallel solid curve (separated by an arrow) denotes the gas cloud after the density increase by a factor 10 due to a dissipative contraction. *Dashed lines*: the total-mass constant lines, assuming a baryon fraction of 0.1.

**Figure 2.** Mass of a self-shielded core against UV photons, $M_c$, as a function of total mass $M_T$ (including dark matter for a baryon fraction of 0.1) for several values of the threshold $\nu$. *Solid lines*: for $J_{-21} = 0.1$. *Dotted lines*: for $J_{-21} = 1.0$. Attached numbers denote the value of $\nu$.

**Figure 3.** Rms density fluctuation $\Delta$ as a function of the top-hat scale $R$ for two filters, top-hat filter (solid line) and Gaussian filter (dotted line). Both curves are normalized so that $\Delta$ is unity at the top-hat lengthscale $16h_{50}^{-1}$Mpc with $h_{50} = 1$.

**Figure 4.** Threshold of density fluctuation, $\nu$, affected by the UV background. (a) The mass-dependency of $\nu$ at $z = 0$ with $b = 2$, for several constraints: *dash-dotted line*: $\nu$ for collapse before $z = 0$; *dotted line*: $\nu$ for cooling before $z = 0$ (without UV); *dashed line*: $\nu$ for forming a self-shielded core against UV photons represented by $J_{-21} = 0.1$; *solid line*: adopted $\nu$ to estimate the mass function. (b) The dependence of $\nu$ on the UV flux $J_{-21}$ at the mass scale $M_b = 10^{11} M_\odot$, with bias parameters $b = 1$ and 2. *Solid lines*: for $z = 0$. *Dotted lines*: for $z = 2$.

**Figure 5.** Number density of objects as a function of baryonic mass $M_b$ at $z = 0$, with a bias $b = 2$, in the Press-Schechter formalism (a) and the peaks formalism (b). *Solid lines*: mass function for collapsed objects at the present epoch. *Dotted lines*: mass function for cooled objects at the present epoch. *Dashed lines*: mass function for cooled and self-shielded cores against UV with $J_{-21} = 0.1$.

**Figure 6.** Time-evolution of the mass function predicted in the Press-Schechter formalism with $b = 2$, at $z = 3$ (*dash-dotted lines*), $z = 2$ (*dashed lines*), $z = 1$ (*dotted lines*), and $z = 0$ (*solid lines*). (a) mass function for all collapsed objects. (b) the case for $J_{-21}(z = 2) = 1.0$

and $J_{-21}(z = 0) = 0.1$. (c) the case for $J_{-21}(z = 2) = 3.0$ and $J_{-21}(z = 0) = 0.1$. (d) the case for $J_{-21}(z = 2) = 3.0$ and $J_{-21}(z = 0) = 0.05$. The adopted evolution of $J_{-21}$ is given in eq.(15).

**Figure 7.** The same as Fig.6 but for the peaks formalism.

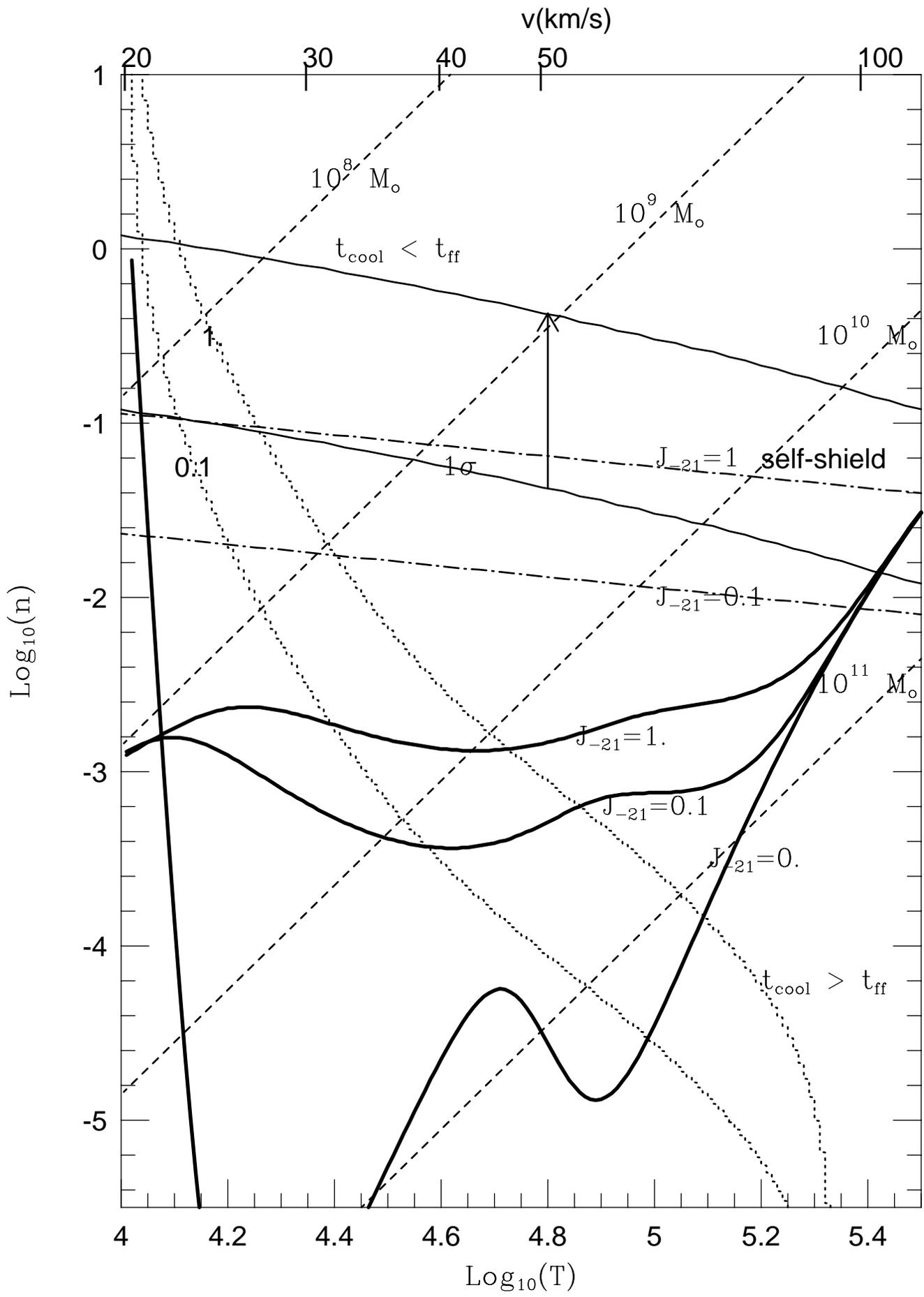

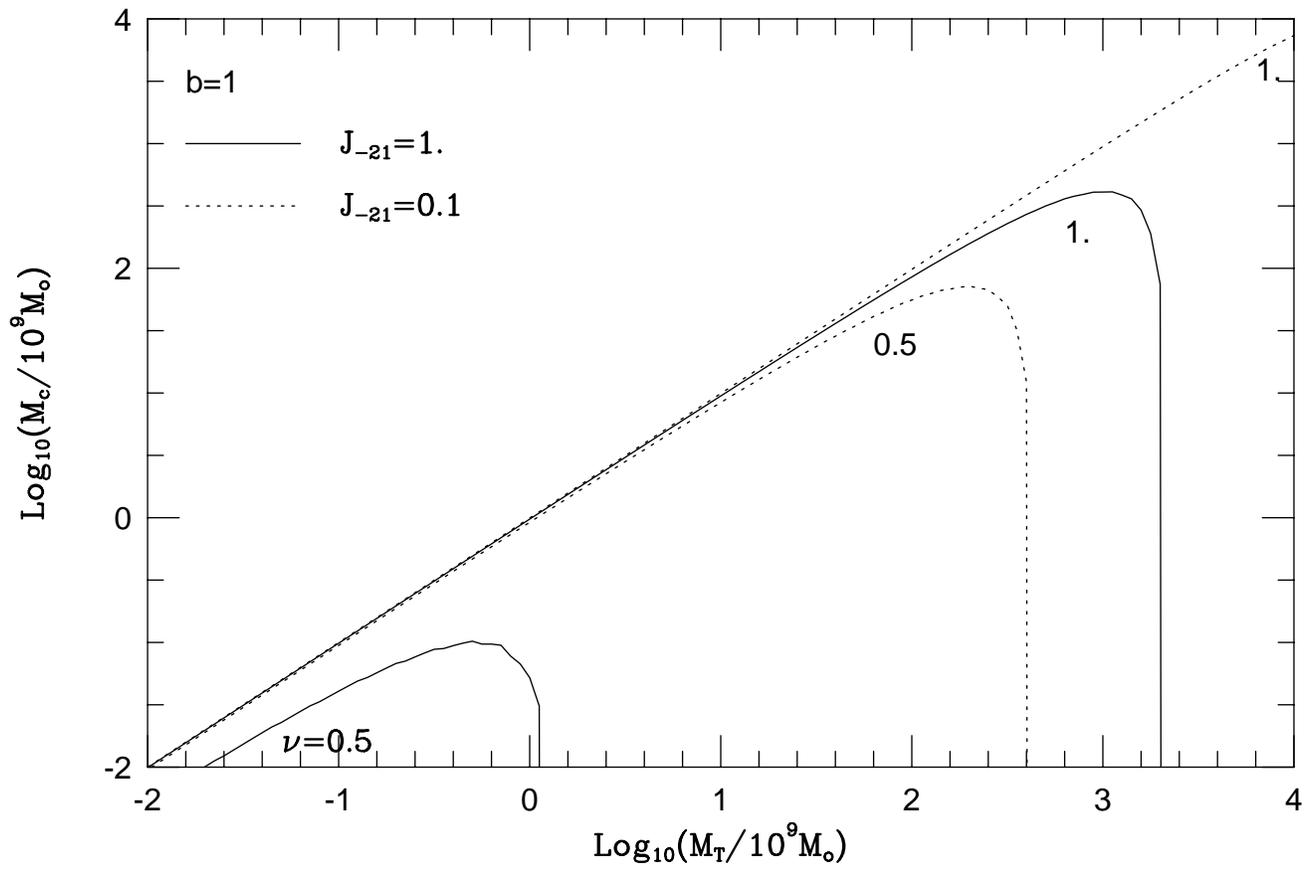

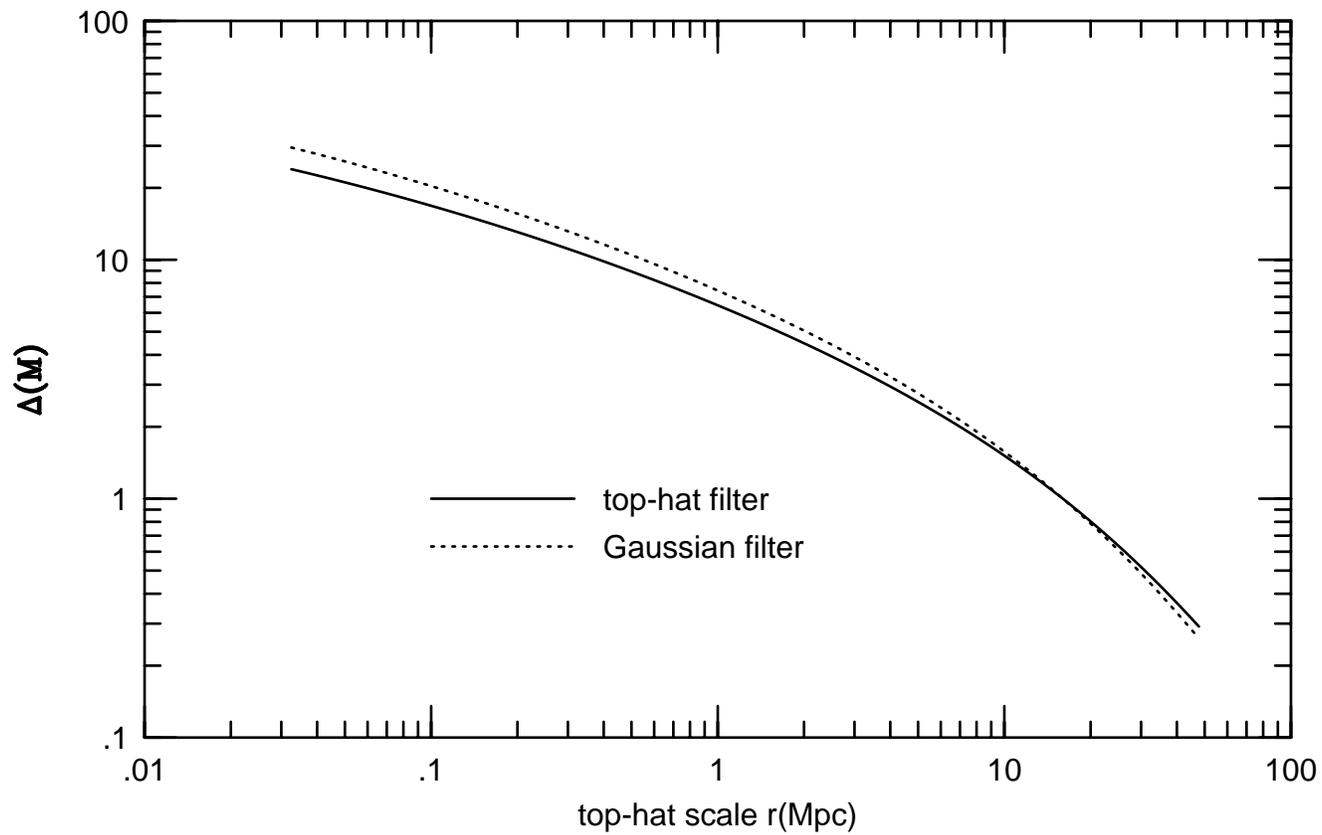

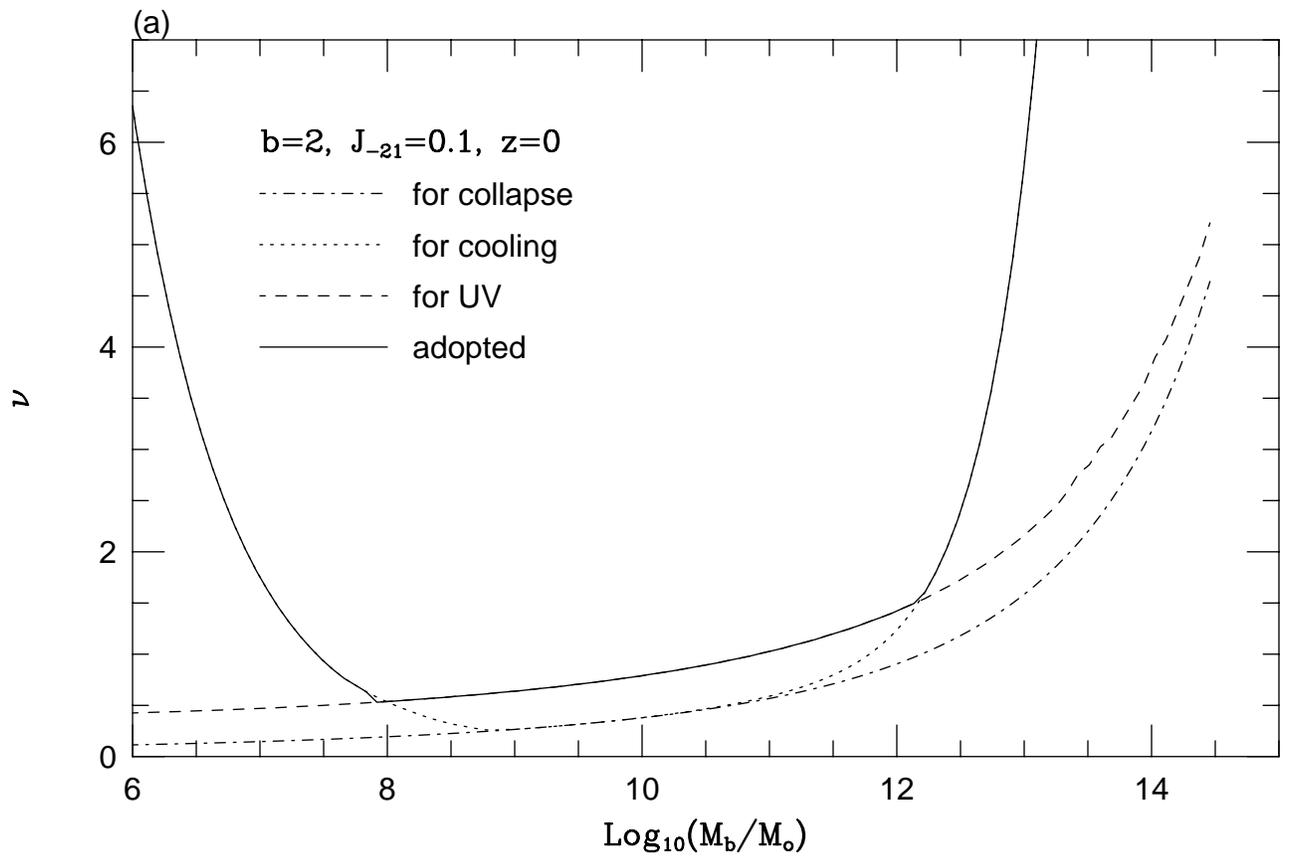

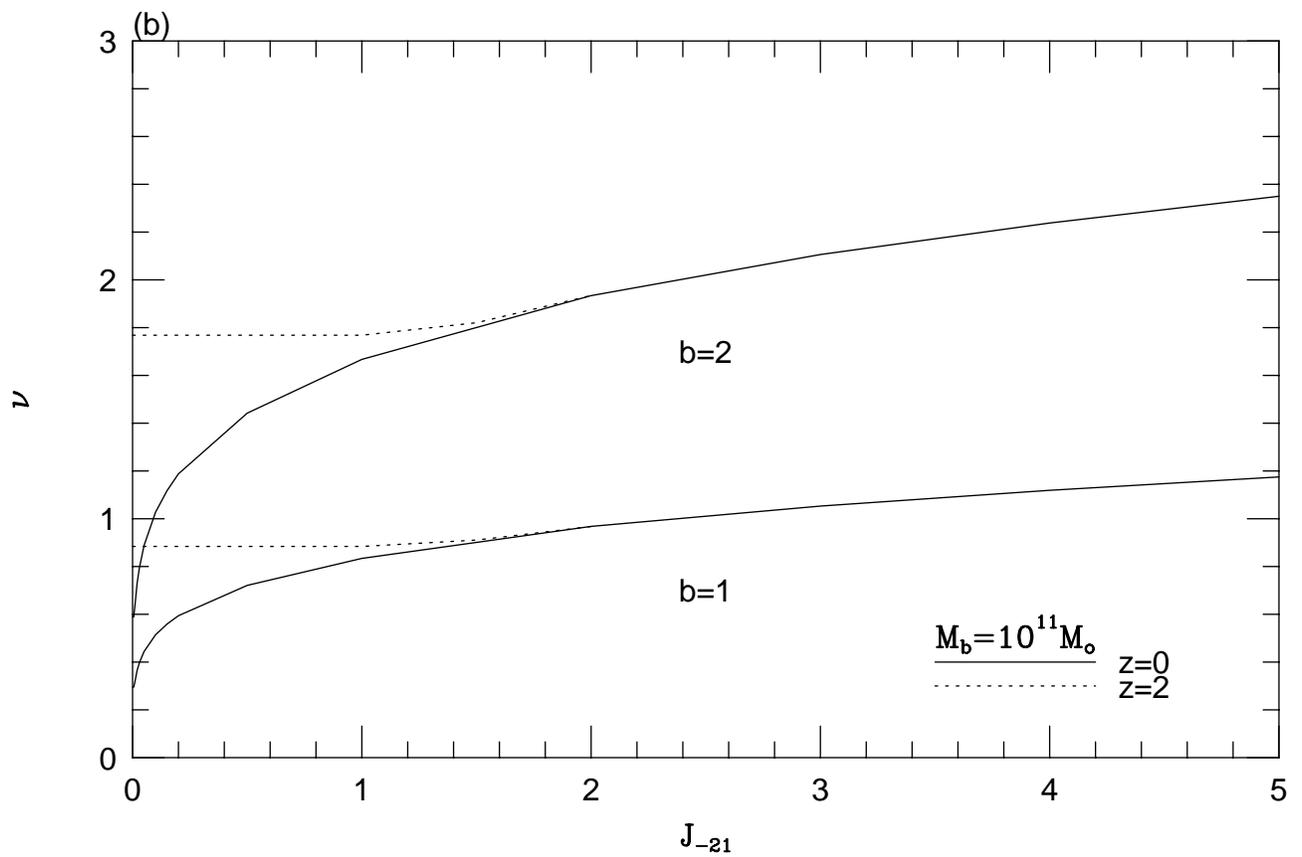

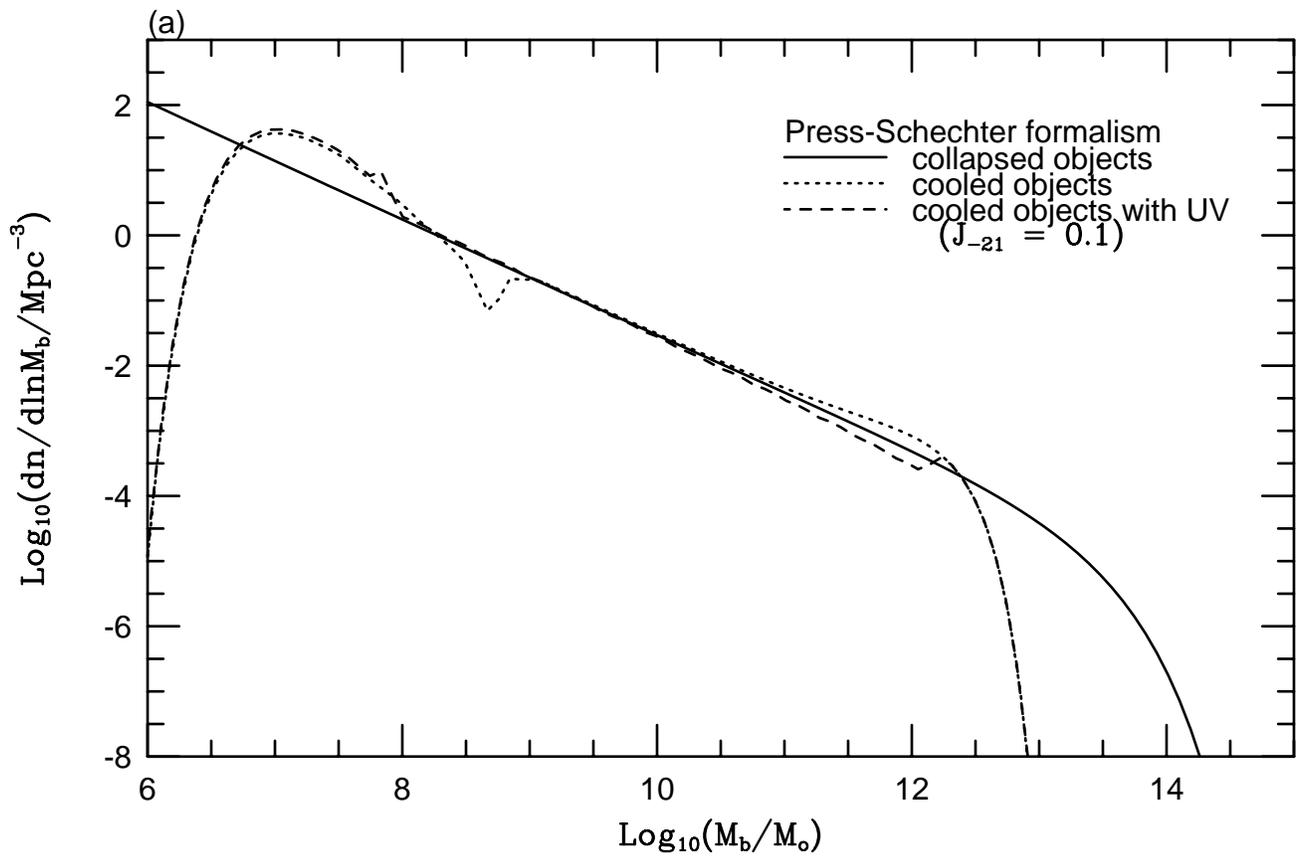

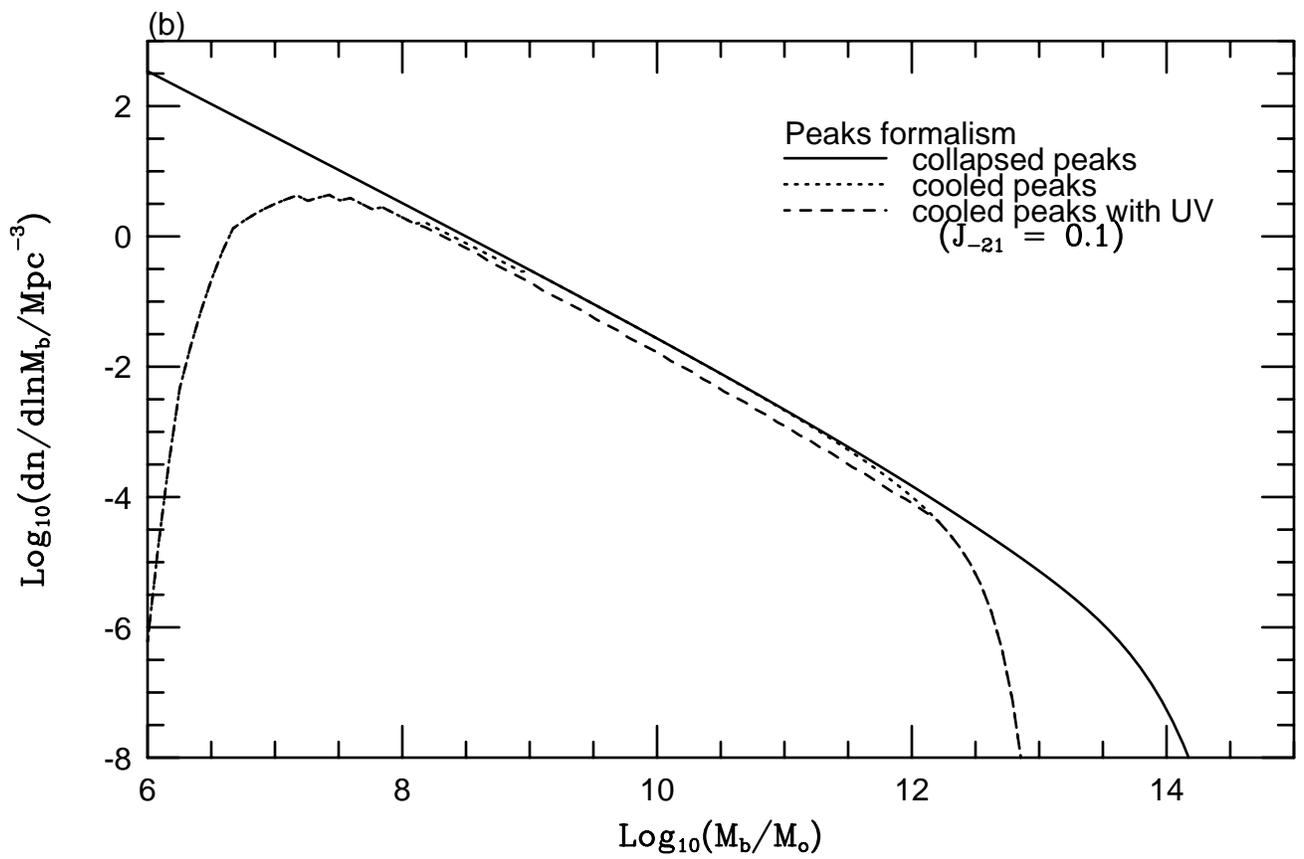

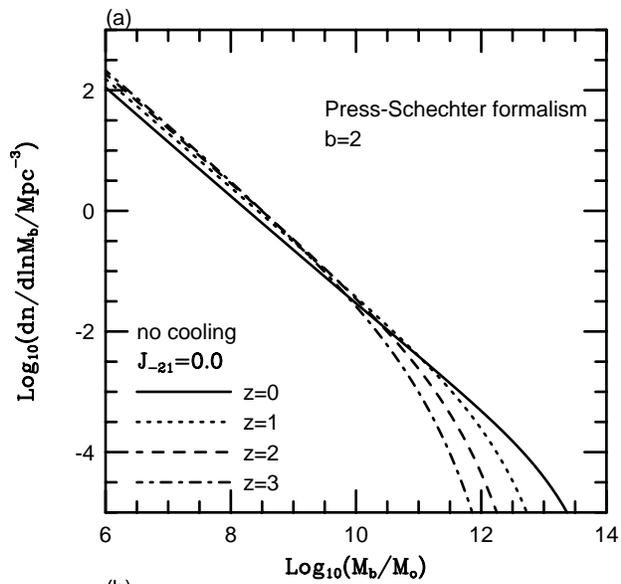
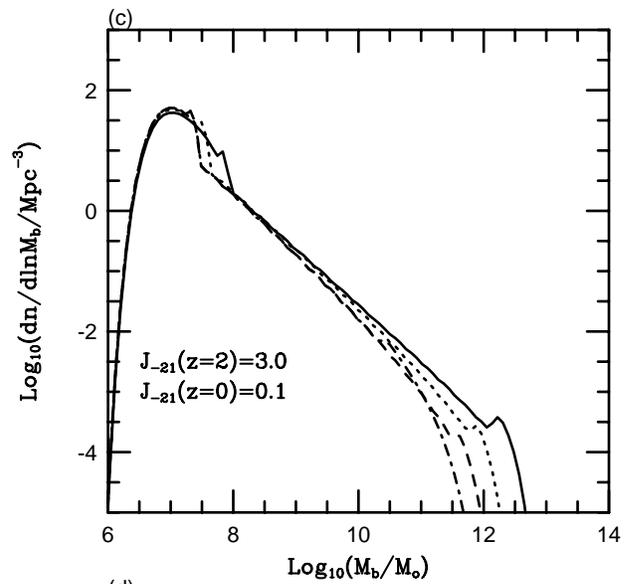
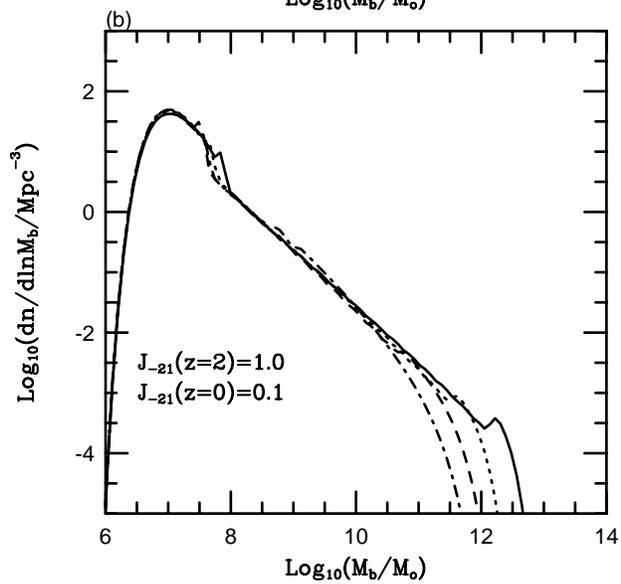
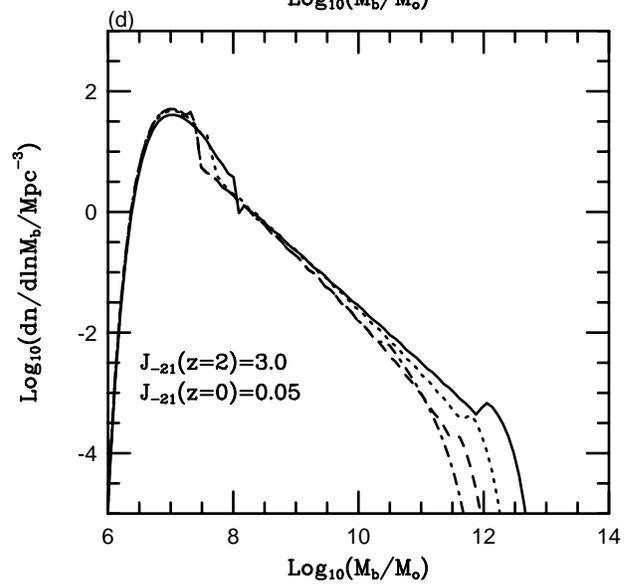

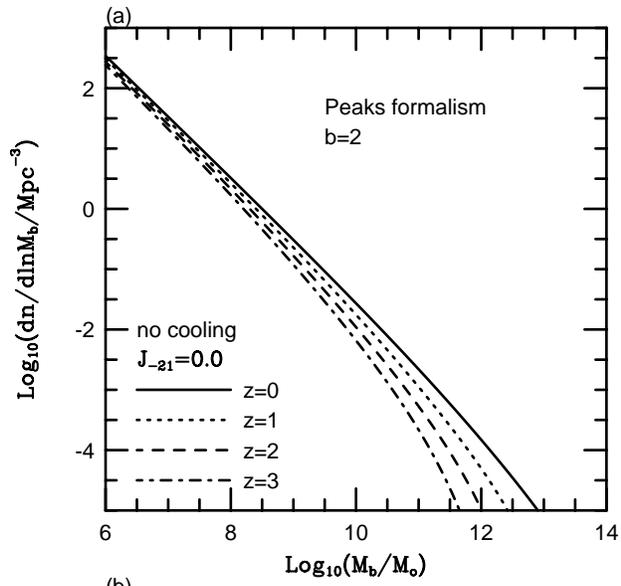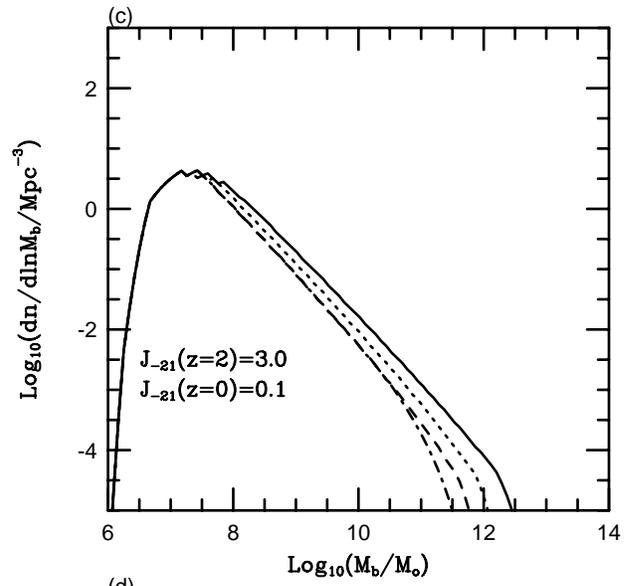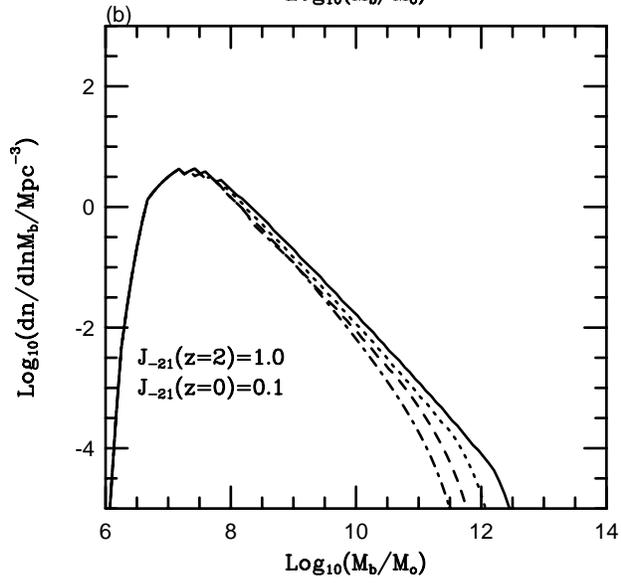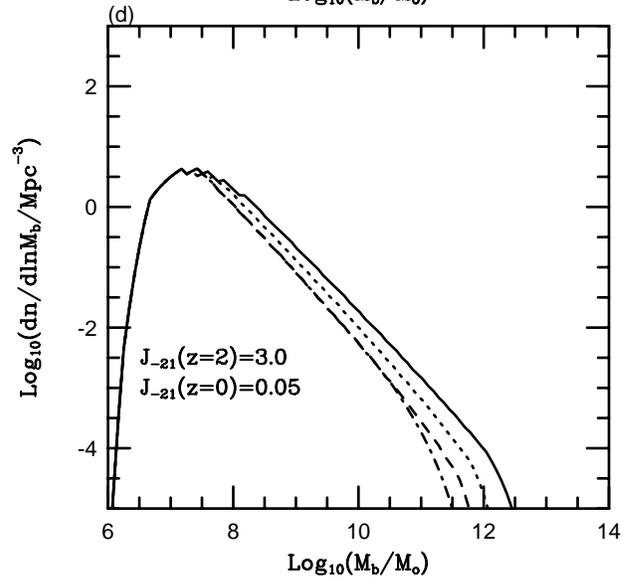